\documentclass[journal,draftclsnofoot,onecolumn,12pt]{IEEEtran}
\usepackage{amsmath,amsfonts}
\usepackage{algorithmic}
\usepackage{algorithm}
\usepackage{array}
\usepackage[caption=false,font=normalsize,labelfont=sf,textfont=sf]{subfig}
\usepackage{textcomp}
\usepackage{stfloats}
\usepackage{url}
\usepackage{verbatim}
\usepackage{graphicx}
\usepackage{cite}
\usepackage{color}
\usepackage{amsthm}
\usepackage{balance}
\usepackage{epstopdf}
\usepackage{latexsym,bm}
\newtheorem{theorem}{Theorem}

\newtheorem{corollary}{Corollary}

\newtheorem{remark}{Remark}\makeatletter
\newlength{\figwidth}
\ifCLASSOPTIONonecolumn
\else
\fi
\begin{document}

\title{Impact of RHIs and ipSIC on Active RIS-NOMA Systems with Low-Precision ADCs}

\author{Qianqian Li,
 Hua Li, Shiya Hao, Lintao Li, Xiaoming Dai
\thanks{The authors are with the School of Computer and Communication Engineering, University of Science and Technology Beijing, Beijing 100083, China, and also with the Shunde Innovation School, University of Science and Technology Beijing, Foshan 528399, China (email: liqianqian\_lf@163.com;  
daixiaoming@ustb.edu.cn).\\
\textit{\emph{Corresponding author: Xiaoming Dai}.}}
}

\markboth{Journal of \LaTeX\ Class Files,~Vol.~14, No.~8, August~2021}%
{Shell \MakeLowercase{\textit{et al.}}: A Sample Article Using IEEEtran.cls for IEEE Journals}
\maketitle

\begin{abstract}
This study evaluates the performance of an active reconfigurable intelligent surface (ARIS)-assisted non-orthogonal multiple access (NOMA) system employing low-precision analog-to-digital converters (ADCs). 
Analytical approximations for the outage probability (OP) are derived, considering residual hardware impairments (RHIs) and imperfect successive interference cancellation (ipSIC). Additionally, we analyze the asymptotic OP, system throughput, and diversity order at high signal-to-noise ratios (SNRs).
Simulation results demonstrate that the proposed quantized ARIS-NOMA system outperforms its passive counterpart (PRIS-NOMA), achieving lower OP and higher throughput with reduced transmit power requirements and fewer reflecting elements. 
Moreover, the outage performance of both quantized ARIS-NOMA and PRIS-NOMA systems demonstrates significant improvement as the number of reflecting elements increases. The negative impacts of low-precision ADCs can be effectively mitigated by optimizing transmit power and scaling the number of reflecting elements.
\end{abstract}

 \vspace{-7pt} 
\section{Introduction}
\IEEEPARstart{N}{on}-orthogonal multiple access (NOMA) has garnered significant interest due to its ability to address multi-user access constraints and improve spectral efficiency\cite{paper_1}.
A key challenge in broadening the applicability of NOMA lies in effectively managing the directionality of user channel vectors.
Recently, reconfigurable intelligent surfaces (RISs) have emerged as a transformative technology. By dynamically adjusting the phases of their reflecting elements, RISs optimize radio propagation environments, thereby improving signal reception and overall network performance \cite{paper_2}.
In this context, passive RIS-assisted NOMA (PRIS-NOMA) systems have been proposed as a cost-effective approach for next-generation wireless networks.
The work in \cite{paper_a3} introduced an optimal power allocation strategy for downlink PRIS-NOMA systems, tailored for short-packet communications. 
To improve energy efficiency, the reflecting and transmitting beamforming was jointly optimized for the PRIS-NOMA systems in \cite{paper_4}.
However, PRIS-NOMA systems are constrained by the double-fading effect, a fundamental limitation that significantly degrades system performance\cite{paper_5}.
To address this challenge, active reconfigurable intelligent surface (ARIS) technology has been developed, integrating reflective amplifiers into its elements to strengthen incident signals and achieve efficient signal amplification with minimal power consumption \cite{paper_6}.
The work in \cite{paper_7a} proposed an alternating optimization algorithm to address the non-convex optimization problem in an uplink ARIS-assisted NOMA (ARIS-NOMA) system. The outage probability (OP) of the ARIS-NOMA system under cascaded Nakagami-$m$ fading channels was derived in \cite{paper_9}. These works demonstrate the potential of ARIS to overcome the shortcomings of PRIS and enhance NOMA performance.
The significant power consumption of high-precision analog-to-digital converters (ADCs) poses a critical challenge in wireless communications, as power requirements escalate exponentially with increasing precision levels. Employing low-precision ADCs offers a promising approach to reduce power consumption in future wireless systems, providing advantages such as cost-effectiveness, energy efficiency, and simplified implementation \cite{paper_13}.
However, practical transceivers are susceptible to hardware impairments that substantially degrade system performance. Although analog and digital signal processing techniques can partially mitigate these impairments, residual hardware impairments (RHIs) remain an inevitable concern \cite{paper_11}. Furthermore, imperfect successive interference cancellation (ipSIC), resulting from instrumentation errors and error propagation, further compromises decoding accuracy \cite{paper_12}.
Prior studies, such as \cite{paper_a3} and \cite{paper_4}, analyzed PRIS-NOMA with ideal hardware, while \cite{paper_7a} explored ARIS-NOMA but neglected the impact of low-precision ADCs and RHIs. Additionally, \cite{paper_9} derived the OP for ARIS-NOMA under Nakagami-$m$ fading without addressing ipSIC and low-precision ADCs. 

In this work, we jointly analyze the effects of RHIs, ipSIC, and low-precision ADCs in ARIS-NOMA systems, deriving analytical and asymptotic OP expressions. The proposed model bridges a key gap in current research, enabling spectral efficiency improvements while maintaining low power and hardware complexity.
We further demonstrate that ARIS-NOMA achieves superior outage performance compared to ARIS-OMA under these practical system constraints.
 \vspace{-5pt} 
\section{System Model}
\begin{figure} [t]
\centering
\includegraphics[width=0.45\textwidth]{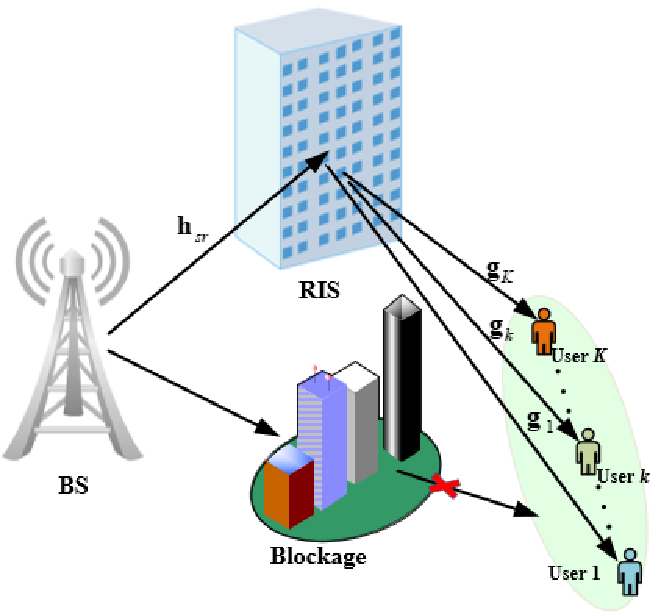}
\caption{A downlink RIS-NOMA system model.}
\label{fig_01}
\end{figure}
We consider a downlink RIS-NOMA system equipped with low-precision ADCs, comprising a base station (BS), a RIS with $M$ reflecting elements, and $K$ single-antenna NOMA users, as depicted in Fig. 1. The small-scale fading coefficient from BS to RIS is denoted by ${{\bf{h}}_{sr}} = {\left[ {{h_{sr,1}},...,{h_{sr,m}},...,{h_{sr,M}}} \right]^H} \in {\mathbb{C}^{M \times 1}}$, and the coefficient from RIS to user $k$, $k \in \left\{ {1,2,...,K} \right\}$, is represented by ${{\bf{g}}_k} = {\left[ {{g_{k,1}},...,{g_{k,m}},...,{g_{k,M}}} \right]^H} \in {\mathbb{C}^{M \times 1}}$. All channels are assumed to experience independent block Rayleigh fading\footnote{ Rayleigh fading is used to model scenarios with significant multipath scattering, and future work will investigate system performance optimization algorithms using Rician or Nakagami-m fading in LoS-dominated scenarios.}
with small-scale fading coefficients distributed according to a Gaussian distribution, i.e., ${\cal C}{\cal N}\left( {0,1} \right)$. The path loss exponent for large-scale fading is denoted by $\alpha$, and the distances from BS to RIS and from RIS to user $k$ are represented by ${d_{sr}}$ and ${d_{k}}$, respectively.
The reflection matrix of the ARIS is defined as ${\bf{\Theta }} = diag\left( {\bm{\theta }} \right)$, where ${\bm{\theta }} = \left[ {{\beta _1}{e^{j{\theta _1}}},...,{\beta _m}{e^{j{\theta _m}}},...,{\beta _M}{e^{j{\theta _M}}}} \right]$,
and ${\theta _m} \in \left[ {0,2\pi } \right)$ represents the phase shift of the $m$-th reflecting element. 
Due to amplifier incorporation in ARIS, the amplification factor satisfies ${\beta _m} > 1$ for all $m$ $\left(1 \le m \le M\right)$.
Unlike PRIS, ARIS introduces non-negligible thermal noise, which is typically neglected in PRIS systems.
To facilitate analytical derivations, we assume identical amplification\footnote{Although idealized, this assumption is widely adopted for analytical tractability.
Offline tests with randomly varying $\beta_m\!\sim\!\mathcal{U}[0,\beta_{\max}]$
show negligible changes in OP trends, confirming it as a reliable baseline.} ${\beta_1}=...={\beta_M}={\beta }$, as adopted in \cite{paper_4,paper_5,paper_6}.
Consequently, ${\bf{\Theta }} = \beta {\bf{\Phi }}$, where ${\bf{\Phi }} = diag\left\{ {{e^{j{\theta _1}}},...,{e^{j{\theta _m}}},...,{e^{j{\theta _M}}}} \right\}$ corresponds to the reflection matrix of a PRIS.

The BS broadcasts a superimposed coded signal $s = \sum\nolimits_{k = 1}^K {\sqrt {{a_k}} {s_k}}$ to $K$ users, where ${s_k}$ denotes the normalized unit-power symbol for user $k$, and ${a_k}$ represents the power allocation coefficient for user $k$, satisfying ${a_1} \ge  \cdot  \cdot  \cdot  \ge {a_k} \ge  \cdot  \cdot  \cdot  \ge {a_K}$ and $\sum\nolimits_{k = 1}^K {{a_k}}  = 1$. 
The received signal at user $k$ in the ARIS-NOMA system, considering RHIs at both the transmitter and receiver, is expressed as
\begin{align}
\label{e_1} \nonumber
y_k^{act} &= \sqrt {d_k^{ - \alpha }d_{sr}^{ - \alpha }} {\bf{g}}_k^H{\bf{\Theta }}{{\bf{h}}_{sr}}\left( {\sqrt {P_s^{act}}} {s}  + \eta _{t,BS}^{act} \right)\\
& + \sqrt {d_k^{ - \alpha }} {\bf{g}}_k^H{\bf{\Theta }}{{\bf{n}}_r} + \eta _{r,k}^{act} + {n_k},
\end{align}
where $P_s^{act}$ denotes the BS transmission power in the ARIS-NOMA system, and ${{\bf{n}}_r} \sim {\cal C}{\cal N}\left( {{\bf{0}},{N_r}{{\bf{I}}_M}} \right)$ represents the thermal noise introduced by the ARIS with noise power ${N_r}$. Additionally, ${n_k}$ denotes the additive white Gaussian noise (AWGN) at user $k$ with ${{{n}}_k} \sim {\cal C}{\cal N}\left( {0,{N_0}} \right)$. 
The RHI-induced distortions at user $k$ and the BS are modeled as ${\bf{\eta }}_{r,k}^{act} \sim {\cal C}{\cal N}\left( 0,{\kappa _{r,k}^2P_s^{act}{\beta ^2}d_k^{ - \alpha }d_{sr}^{ - \alpha } {{\left| {{\bf{g}}_k^H{\bf{\Phi }}{{\bf{h}}_{sr}}} \right|}^2}} \right)$ and ${\eta _{t,BS}^{act}} \sim {\cal C}{\cal N}\left(0, {\kappa _{t,BS}^2P_s^{act}} \right)$, where $\kappa_{r,k}$ and $\kappa_{t,BS}$ quantify the severity of RHIs.

The additive quantization noise model (AQNM), given by $r_k^{act} ={\lambda _q}{ y_k^{act}}+ n_q^{act}$, can be used to model and evaluate quantization distortion. ${\lambda _q}$ denotes the distortion factor, dependent on the quantization level determined by the $b$-bit, and $n_q^{act}$ represents the quantization noise \cite{paper_a0}.  
The quantized received signal at user $k$ is expressed as
\begin{align}
\label{e_2} \nonumber
r_k^{act}&{\rm{ = }}{\lambda _q}\left( {\sqrt {d_k^{ - \alpha }d_{sr}^{ - \alpha }} {\bf{g}}_k^H{\bf{\Theta }}{{\bf{h}}_{sr}}\left( {\sqrt {P_s^{act}}} {s} + \eta _{t,BS}^{act}\right)} \right.\\
&\left. { + \sqrt {d_k^{ - \alpha }} {\bf{g}}_k^H{\bf{\Theta }}{{\bf{n}}_r} + \eta _{r,k}^{act} + {n_k}} \right) + n_q^{act},
\end{align}
where 
the variance of $n_q^{act}$ is approximated as
\begin{align}
\label{e_3} \nonumber
R_{{n_q}}^{act}& \approx {\lambda _q}\left( {1 - {\lambda _q}} \right)\left[ {d_k^{ - \alpha }d_{sr}^{ - \alpha }{\beta ^2}{{\left| {{\bf{g}}_k^H{\bf{\Phi }}{{\bf{h}}_{sr}}} \right|}^2}} \right.\\
&\left. { \times P_s^{act}\left( {1 + \kappa _{t,BS}^2 + \kappa _{r,k}^2} \right) + d_k^{ - \alpha }{\beta ^2}{N_r}{{\left\| {{\bf{g}}_k^H{\bf{\Phi }}} \right\|^2}}} \right].
\end{align}
The signal-to-interference-plus-noise ratio (SINR) for user $k$ decoding the signal of user $j$ $\left( {1 \le j \le k - 1} \right)$ is given by
\begin{align}
\label{e_4}
\! \!\!\gamma _{k \to j}^{act}\! = \!\!\frac{{\lambda _q^2{\beta ^2}{{\left| {{\bf{g}}_k^H{\bf{\Phi }}{{\bf{h}}_{sr}}} \right|}^2}\rho _s^{act}{a_j}}}{{{\beta ^2}{{\left| {{\bf{g}}_k^H{\bf{\Phi }}{{\bf{h}}_{sr}}} \right|}^2}\!\rho _s^{act}{\vartheta _j} \!\!+\! \!{\varphi _1} \!\!+ \! d_{sr}^\alpha d_k^\alpha \!\left(\! {\lambda _q^2 \!+ \!\epsilon \rho _s^{act}{{\left| {{h_I}} \right|}^2}} \!\right)}},\!
\end{align}
where $\rho _s^{act} = {{P_s^{act}} \mathord{\left/
 {\vphantom {{P_s^{act}} {{N_0}}}} \right.
 \kern-\nulldelimiterspace} {{N_0}}}$ represents the transmit signal-to-noise ratios (SNRs),  ${\vartheta _j} = \lambda _q^2\left( {\sum\nolimits_{i = j + 1}^K {{a_i}}  - 1} \right) + {\lambda _q}\left( {1 + \kappa _{t,BS}^2 + \kappa _{r,k}^2} \right)$, ${\varphi _1} =d_{sr}^\alpha {\beta ^2}{N_r}{\left\| {{\bf{g}}_k^H{\bf{\Phi }}} \right\|^2}{\Delta _1}$, ${\Delta _1} = {{{\lambda _q}} \mathord{\left/
 {\vphantom {{{\lambda _q}} {{N_0}}}} \right.
 \kern-\nulldelimiterspace} {{N_0}}}$,
 and ${h_I} \sim {\cal C}{\cal N}\left( {0,{\Omega _I}} \right)$ represents the corresponding complex coefficient of the residual interference during the ipSIC operation \cite{paper_12}.
 The conditions $\epsilon=0$ and $\epsilon \in \left( {0,1} \right] $ correspond to pSIC and ipSIC, respectively. 
 The SINR for user $k$ decoding its own signal is given by
\begin{align}
\label{e_5}
\!\!\!\gamma _k^{act}\!\! =\!\! \frac{{\lambda _q^2{\beta ^2}{{\left| {{\bf{g}}_k^H{\bf{\Phi }}{{\bf{h}}_{sr}}} \right|}^2}\rho _s^{act}{a_k}}}{{{\beta ^2}{{\left| {{\bf{g}}_k^H{\bf{\Phi }}{{\bf{h}}_{sr}}} \right|}^2}\!\rho _s^{act}{\vartheta _k}\!\! +\! {\varphi _1}\!\! + \! d_{sr}^\alpha d_k^\alpha \!\left( \!{\lambda _q^2 \!+\! \epsilon \rho _s^{act}{{\left| {{h_I}} \right|}^2}} \!\right)}},\!
\end{align}
where ${\vartheta _k} = \lambda _q^2\left( {\sum\nolimits_{i = k + 1}^K {{a_i}}  - 1} \right) + {\lambda _q}\left( {1 + \kappa _{t,BS}^2 + \kappa _{r,k}^2} \right)$.
After removing the interference from the previous $K-1$ users, the SINR for user $K$ decoding its own signal is given by
\begin{align}
\label{e_6}
\!\!\!\gamma _K^{act} \!\!=\!\! \frac{{\lambda _q^2{\beta ^2}{{\left| {{\bf{g}}_K^H{\bf{\Phi }}{{\bf{h}}_{sr}}} \right|}^2}\rho _s^{act}{a_K}}}{{{\beta ^2}{{\left| {{\bf{g}}_K^H{\bf{\Phi }}{{\bf{h}}_{sr}}} \right|}^2}\!\rho _s^{act}{\vartheta _K} \!\!+\! {\varphi _1}\! \!+\! d_{sr}^\alpha d_K^\alpha\! \left(\! {\lambda _q^2\! +\! \epsilon \rho _s^{act}{{\left| {{h_I}} \right|}^2}} \!\right)}},\!
\end{align}
where ${\vartheta _K} =  - \lambda _q^2 + {\lambda _q}\left( {1 + \kappa _{t,BS}^2 + \kappa _{r,K}^2} \right)$.
\section{Performance Analysis}
We define ${\left(\xi _M\right)}^{2}\!\!\!=\!\!{\left| {{\bf{g}}_k^H{\bf{\Phi }}{{\bf{h}}_{sr}}} \right|^2} \!\!\!= \!\!{\left| {\sum\nolimits_{m = 1}^M {{g_{k,m}}} {e^{j{\theta _m}}}{h_{sr,m}}} \right|^2}$.
Note that $\xi _M^{}= { {\sum\nolimits_{m = 1}^M {\left| {{g_{k,m}}} \right|} \left| {{h_{sr,m}}} \right|}} = \sum\nolimits_{m = 1}^M {\xi _m^{}}$ is derived with the assistance of coherent phase shifting, which maximizes the received performance for the target users.
The cumulative distribution function (CDF) of $\xi _M^{}$ is approximated as
\cite{paper_15}
\begin{equation}
\label{e_7}
{F_{\xi _{M}}}\left( x \right) \approx  \frac{1}{{\Gamma \left( {{\mu _0} + 1} \right)}}\gamma \left( {{\mu _0} + 1,\frac{x}{{{\varphi _0}}}} \right),
\end{equation}
where ${\mu _0} = \frac{{\left( {M + 1} \right){\pi ^2} - 16}}{{16 - {\pi ^2}}}$ and ${\varphi _0} = \frac{{16 - {\pi ^2}}}{{4\pi }}$ represent the mean and variance of ${\xi _{M}}$, respectively. $\Gamma \left(  \cdot  \right)$ denotes the Gamma function and $\gamma \left( { \cdot , \cdot } \right)$ is the lower incomplete Gamma function.
\vspace{-6pt} 
\subsection{OP Analysis}
 An outage event for user $k$ occurs when user $k$ cannot correctly decode $s_k$.
We define ${\Lambda _{k,j}} \buildrel \Delta \over = \left\{ {\gamma _{k \to j}^{} \le {\gamma _{thj}}} \right\}$ as the event in which user $k$ fails to decode the signal of user $j$ $\left( {1 \le j \le k} \right)$, with ${\bar \Lambda _{k,j}}$ denoting its complementary event. Consequently, the OP of user $k$ is expressed as
\begin{equation}
\label{e_8}
P_{k}^{} = 1 - \Pr \left( {{{\bar \Lambda }_{k,1}}  \cap ... \cap {{\bar \Lambda }_{k,k}} } \right),
\end{equation}
where ${\gamma _{thj}} = {2^{{R_j}}} - 1$ denotes the outage threshold for user $k$, and ${R_j}$ is the target rate for decoding $s_j$.
\begin{theorem}
\label{theorem_1}
Under the ipSIC condition (i.e., $\epsilon \in \left( {0,1} \right] $), the analytical approximation for the OP of user $k$ in the quantized ARIS-NOMA system with RHIs is given by
\begin{align}
\label{e_9} \nonumber
&P_{act,k}^{ipSIC}=\frac{1}{{\Gamma \left( {{\mu _0} + 1} \right)}}\sum\limits_{p = 1}^P {{A_p}} \gamma \left( {{\mu _0} + 1,} \right.\\
&\left. {\frac{1}{{{\varphi _0}}}\sqrt {\varphi _{}^{act*}d_{sr}^\alpha \left( {{\beta ^2}M{N_r}{\Delta _1} + d_k^\alpha \left( {\lambda _q^2 + \rho _s^{act}\epsilon {\Omega _I}{z_p}} \right)} \right)} } \right),
\end{align}
where ${A_p}{\rm{ = }}\frac{{{{\left( {P!} \right)}^2}{x_p}}}{{{{\left( {{L_{P + 1}}\left( {{x_p}} \right)} \right)}^2}}}$ is the $p$-th weight for Gauss-Laguerre integration, ${x_p}$ is the $p$-th zero point of Laguerre polynomial $L_P(x)=\frac{e^x}{P!} \frac{d^P}{d x^P}\left(e^{-x} x^P\right)$, and $P$ is a tradeoff parameter balancing complexity and accuracy. ${\varphi ^{act*}} \!\!=\!\! \max \left[ {\varphi _1^{act},\varphi _2^{act},...,\varphi _k^{act}} \right]$, $\varphi _k^{act} = \frac{{{\gamma _{thk}}}}{{{\beta ^2}\rho _s^{act}\left( {\lambda _q^2{a_k} - {\gamma _{thk}}{\vartheta _k}} \right)}}$ with $\lambda _q^2{a_k} > {\gamma _{thk}}{\vartheta _k}$.

\begin{proof}
Please see Appendix A.
\end{proof}
\end{theorem}
\begin{theorem}
\label{theorem_2}
Under the ipSIC condition, the analytical approximation for the OP of user $k$ in the quantized PRIS-NOMA system with RHIs is given by
\begin{align}
\label{e_12} \nonumber
&P_{pas,k}^{ipSIC} = \frac{1}{{\Gamma \left( {{\mu _0} + 1} \right)}}\sum\limits_{p = 1}^P {{B_p}}  \times \\
&\gamma \left( {{\mu _0} + 1,\frac{1}{{{\varphi _0}}}\sqrt {\varphi _{}^{pas*}d_{sr}^\alpha d_k^\alpha \left( {\lambda _q^2 + \rho _s^{pas}\epsilon{\Omega _I}{z_p}} \right)} } \right),
\end{align}
where $\rho _s^{pas} = {{P_s^{pas}} \mathord{\left/
 {\vphantom {{P_s^{pas}} {{N_0}}}} \right.
 \kern-\nulldelimiterspace} {{N_0}}}$, $ {P_s^{pas}}$ denotes the BS transmission power in the PRIS-NOMA system, $B_p$ is the p-th weight for Gauss-Laguerre integration, and $z_p$ is the p-th zero point of Laguerre polynomial. ${\varphi ^{pas*}} = \max \left[ {\varphi _1^{pas},\varphi _2^{pas},...,\varphi _k^{pas}} \right]$, $\varphi _k^{pas} = \frac{{{\gamma _{thk}}}}{{\rho _s^{pas}\left( {\lambda _q^2{a_k} - {\gamma _{thk}}{\vartheta _k}} \right)}}$, with $\lambda _q^2{a_k} > {\gamma _{thk}}{\vartheta _k}$.

 \begin{proof}
Please see Appendix B.
\end{proof}
\end{theorem}
\begin{corollary}
\label{corollary_1}
Under the pSIC condition (i.e., ${\epsilon = 0}$), the OP of user $k$ in the quantized ARIS-NOMA system with RHIs is given by
\begin{align}
\label{e_14} 
P_{act,k}^{pSIC} \!\!=  \!\!\frac{1}{{\Gamma \!\left(\! {{\mu _0} \!+\! 1} \right)}}\!\gamma\! \left(\! {{\mu _0}\! + \!1,} {\frac{1}{{{\varphi _0}}}\!\sqrt {\!\varphi _{}^{act*}\!d_{sr}^\alpha\! \left( \!{{\beta ^2}M\!{N_r}{\Delta _1}\! \!+\! d_k^\alpha \lambda _q^2} \right)} } \!\right)\!.
\end{align}
\end{corollary}
\begin{corollary}
\label{corollary_2}
Under the pSIC condition, the OP of user $k$ in the quantized PRIS-NOMA system with RHIs is given by
\begin{align}
\label{e_16}
P_{pas,k}^{pSIC} \!= \!\frac{1}{{\Gamma \left( {{\mu _0}\! +\! 1} \right)}}\gamma \!\left( {{\mu _0} \!+\! 1,\frac{1}{{{\varphi _0}}}\sqrt {\varphi _{}^{pas*}d_{sr}^\alpha d_k^\alpha \lambda _q^2} } \right).
\end{align}
\end{corollary}
\vspace{-5pt} 
\subsection{Diversity Order Analysis}
The diversity order of user $k$ for the quantized ARIS-NOMA and PRIS-NOMA systems is defined as
\begin{align}
\label{e_18}
{d_k} =  - \mathop {\lim }\limits_{{\rho _s} \to \infty } \frac{{\log \left( {P_{k}^\infty \left( {{\rho _s}} \right)} \right)}}{{\log {\rho _s}}},
\end{align}
where ${P_{k}^\infty \left( {{\rho _s}} \right)}$ represents the asymptotic OP of user $k$ in the high SNR regions.
\begin{corollary}
\label{corollary_3}
As $\rho _s^{act} \to \infty $, the asymptotic OP of user $k$ for the quantized ARIS-NOMA systems with RHIs and ipSIC is given by
 \begin{align}
 \label{e_19} \nonumber
&P_{act,k}^{\infty ,ipSIC} = \frac{1}{{\Gamma \left( {{\mu _0} + 1} \right)}}\sum\limits_{p = 1}^P {{A_p}}  \times \\
&\gamma \left( {{\mu _0} + 1,\frac{1}{{{\varphi _0}}}\left. {\sqrt {{\varphi ^{act*}}d_{sr}^\alpha d_k^\alpha \epsilon \rho _s^{act}{\Omega _I}{z_p}} } \right)} \right..
\end{align}
The asymptotic OP of user $k$ in the quantized ARIS-NOMA systems with RHIs and pSIC is given by
 \begin{align}
 \label{e_20} 
P_{act,k}^{\infty ,pSIC}\!=\!\frac{{\sqrt {{{\!\left( {\varphi _{}^{act*}d_{sr}^\alpha } \right)}^{{\mu _0} \!+ \!1}}} }}{{\Gamma\! \left( {{\mu _0} \!+\! 2} \right)\varphi _0^{{\mu _0} + 1}}}{\left( {{\beta ^2}M{N_r}{\Delta _1} \!+\! d_k^\alpha \lambda _q^2} \right)^{\frac{{{\mu _0}\! +\! 1}}{2}}}.
\end{align}
\end{corollary}
\begin{corollary}
\label{corollary_4}
\textit{The asymptotic OP of user $k$ in the quantized PRIS-NOMA systems with RHIs and ipSIC is given by}
 \begin{align}
 \label{e_21}
 P_{pas,k}^{\infty ,ipSIC}\!\!= \!\frac{1}{{\Gamma\! \left( {{\mu _0} \!+ \!1} \right)}}\!\sum\limits_{p = 1}^P  \!{B_p}\gamma \!\left( \!{{\mu _0} \!+\! 1,{\frac{1}{{{\varphi _0}}}\!\sqrt {{\varphi ^{pas*}} \epsilon \rho _s^{pas}{\Omega _I}{z_p}} } } \right).
\end{align}
The asymptotic OP of user $k$ in the quantized PRIS-NOMA systems with RHIs and pSIC is given by
 \begin{align}
 \label{e_22}
P_{pas,k}^{\infty ,pSIC} = \frac{1}{{\Gamma \left( {{\mu _0} + 2} \right)\varphi _0^{{\mu _0} + 1}}}{\left( {{\varphi ^{pas*}}d_{sr}^\alpha d_k^\alpha \lambda _q^2} \right)^{\frac{{{\mu _0} + 1}}{2}}}.
\end{align}
\end{corollary}
 \vspace{-3pt} 
\begin{remark}
\label{remark_1}
Substituting \eqref{e_19} and \eqref{e_20} into \eqref{e_18}, it is found that under ipSIC and pSIC conditions, the diversity orders of the quantized ARIS-NOMA systems with RHIs are 0 and $\frac{{{\mu _0} + 1}}{2}$, respectively. A similar conclusion holds for the quantized PRIS-NOMA system. A zero diversity order implies that the OP converges to an error floor at high SNR due to residual interference caused by ipSIC. It restricts the system's capacity to exploit fading diversity, degrading channel reliability.
\end{remark}
 \vspace{-15pt} 
\subsection{Throughput Analysis}
The throughput of user $k$
is given by
 \begin{align}
 \label{e_23}
T_{\xi ,k}^{\upsilon } = \left( {1 - P_{\xi,k}^{ \upsilon }} \right){R_k},
\end{align}
where $\xi  \in \left\{ {pas,act} \right\}$ and $\upsilon  \in \left\{ {ipSIC,pSIC} \right\}$.
 \vspace{-3pt} 
\section{Numerical Results}
\begin{figure*}[htbp]
\centering
\subfloat[]{\includegraphics[width=0.45\textwidth]{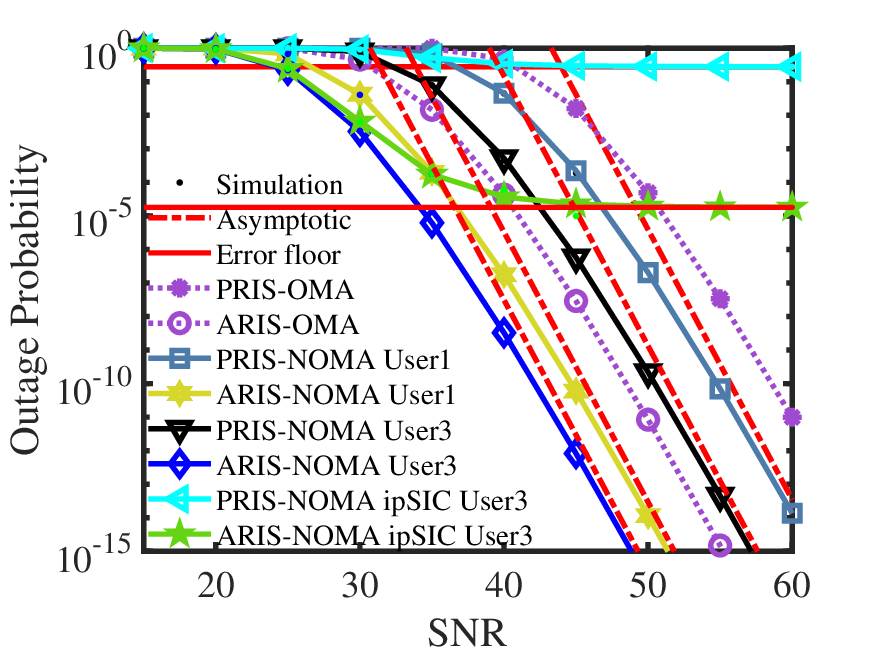}}
\label{fig_1a}
\hfil
\subfloat[]{\includegraphics[width=0.47\textwidth]{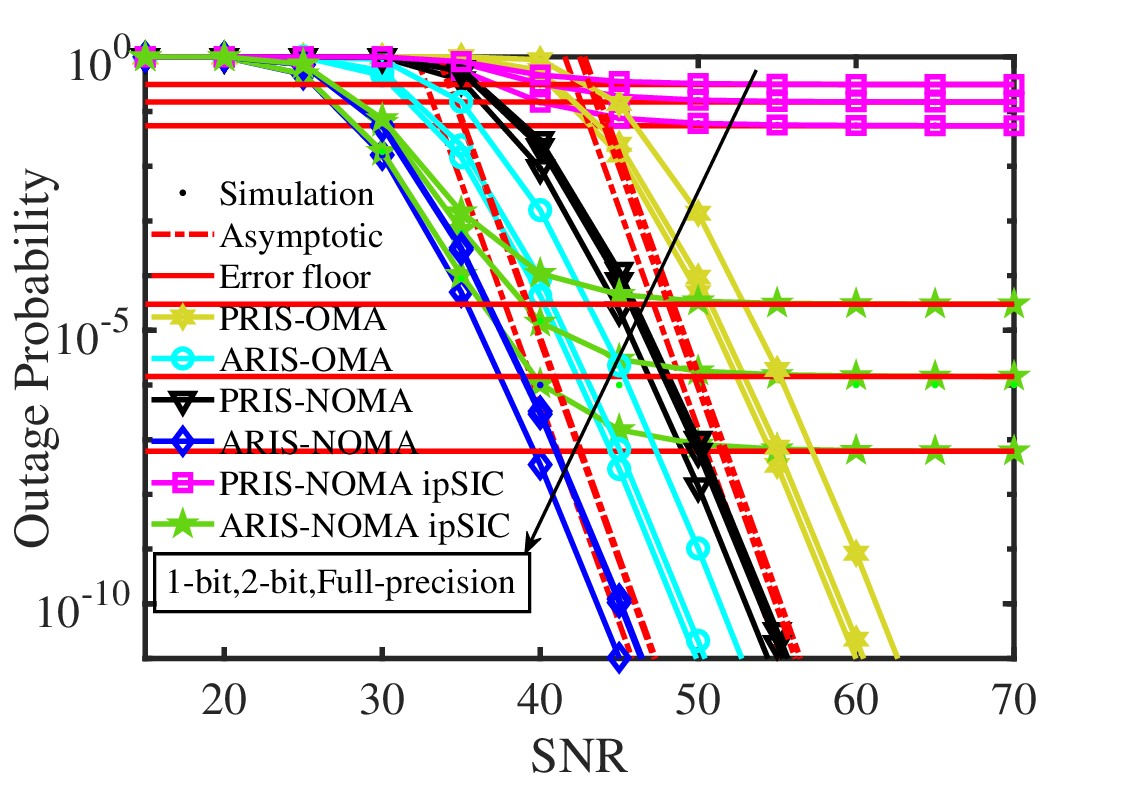}}%
\label{fig_1b}
\caption{(a) OP in full-precision ADCs scenarios versus SNR with $\beta  = 7$. (b) Impact of low-precision ADCs on the OP of user $2$ with $\beta  = 7$.}
\label{fig_1}
\end{figure*}
\begin{figure*}[htbp]
\centering
\subfloat[]{\includegraphics[width=0.45\textwidth]{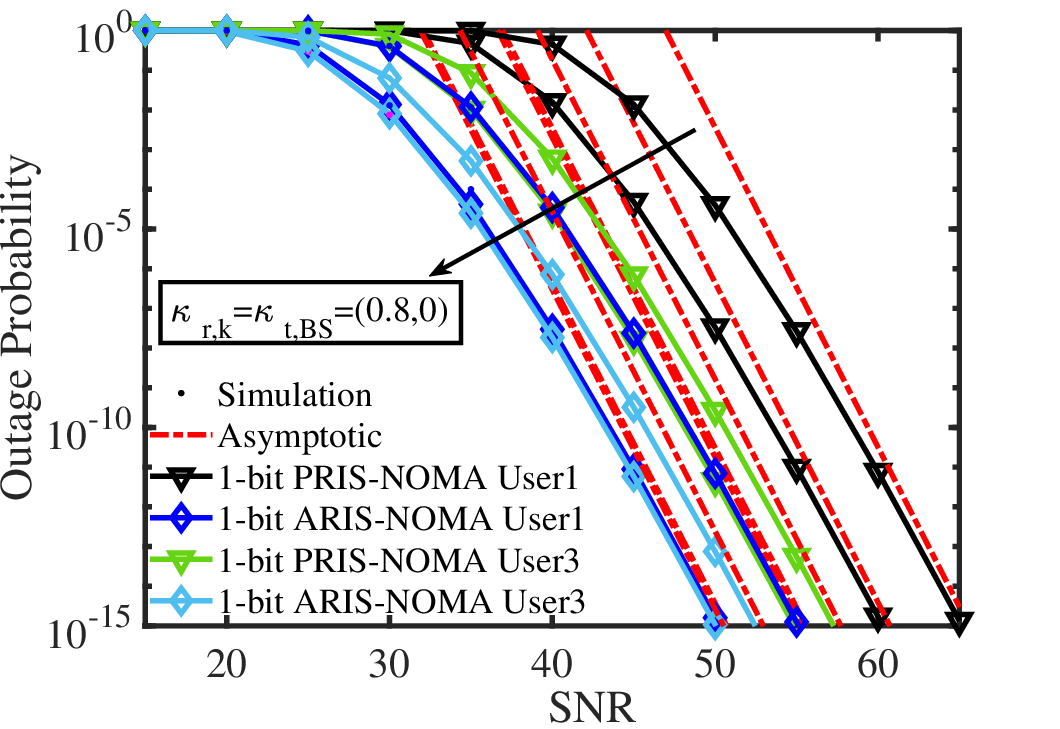}}%
\label{fig_2a}
\hfil
\subfloat[]{\includegraphics[width=0.45\textwidth]{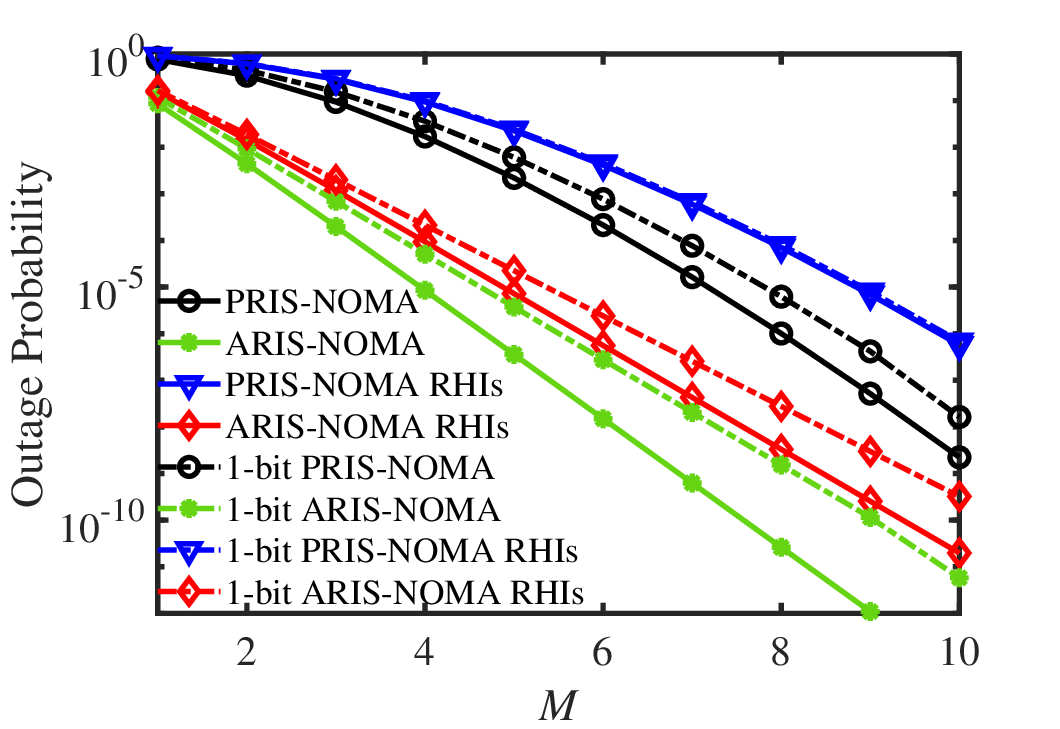}}%
\label{fig_2b}
\caption{(a) Impact of RHIs on OP. (b) OP of user $3$ in ARIS-NOMA and PRIS-NOMA systems versus $M$ at $SNR = 45$dB.}
\label{fig_2}
\end{figure*}
\begin{figure*}[htbp]
\centering
\subfloat[]{\includegraphics[width=0.45\textwidth]{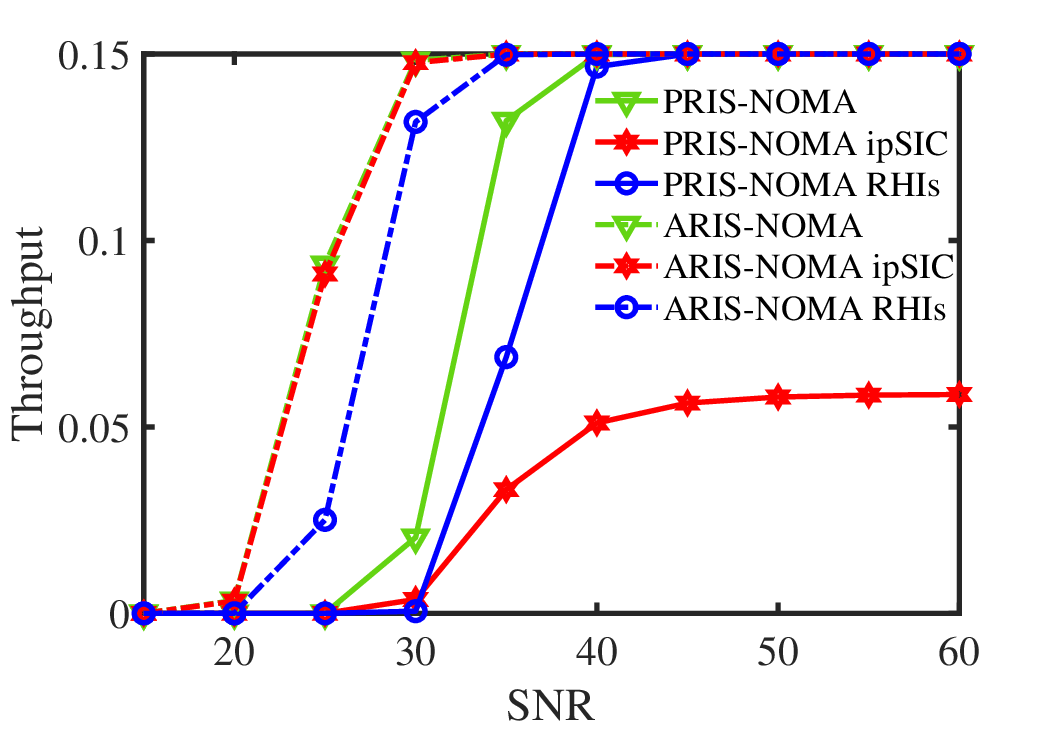}}%
\label{fig_3a}
\hfil
\subfloat[]{\includegraphics[width=0.45\textwidth]{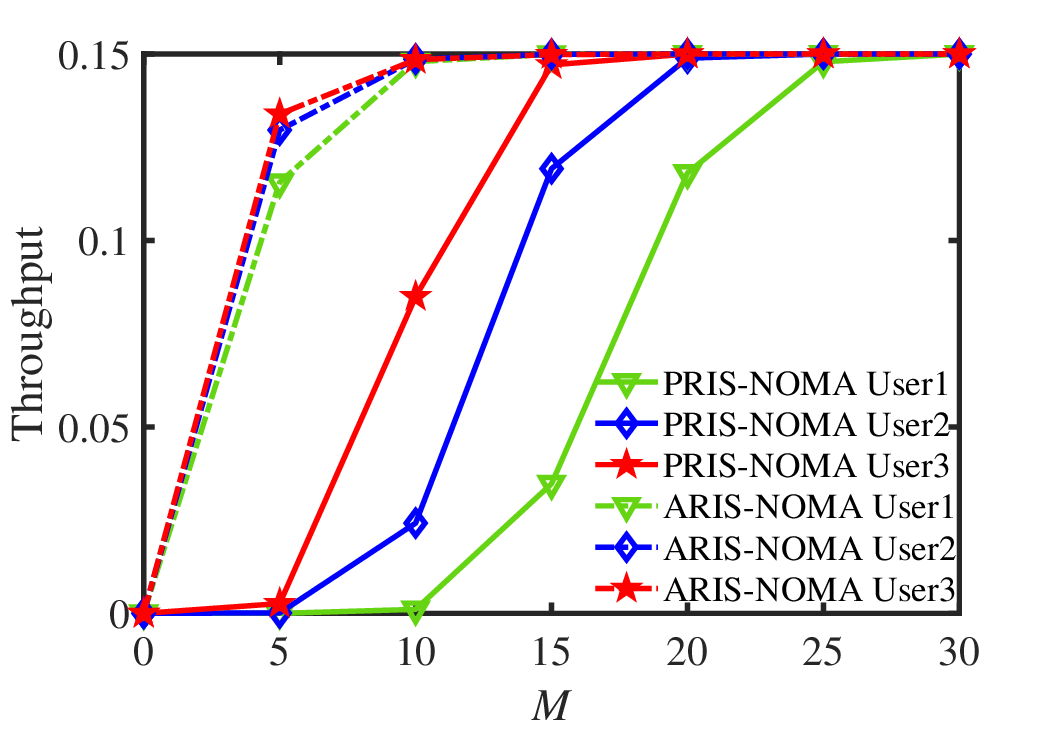}}%
\label{fig_3b}
\caption{(a) Throughput of user $2$ in a 1-bit ADC scenario versus SNR. (b) Throughput in a 1-bit ADC scenario versus $M$.}
\label{fig_3}
\end{figure*}

This section presents numerical results to validate the analytical conclusions for quantized ARIS-NOMA systems.
 The system parameters are set as follows: $\epsilon=0.05$, $\kappa _{t,BS}=\kappa _{r,k}=0.8$, $ M=10$, ${a_1} = 0.45$, ${a_2} = 0.30$, ${a_3} = 0.25$, $\alpha  = 2.2$, and ${R_1} = {R_2} = {R_3} = 0.15$ bit per channel use (BPCU).
The spatial coordinates of BS, RIS, user 3, user
2, and user 1 are set to $(0, 0)$ m, $(10, 5)$ m, $(40, -15)$ m, $(32, 0)$ m, and $(25, 10)$ m, respectively. For fair comparison, we ensure $P_T^{\text{act}} = P_T^{\text{pas}}$, where $P_T^{\text{act}} = P_S^{\text{act}} + P_{\text{RIS}}^{\text{act}} + M(P_{\text{SW}} + P_{\text{DC}})$ and $P_T^{\text{pas}} = P_S^{\text{pas}} + M P_{\text{SW}}$ represent the total power consumption of ARIS-NOMA and PRIS-NOMA systems, respectively. Here, ${P_{\text{SW}}}$ is the power induced by phase-shift switching and control circuits in each element, $P_{\text{RIS}}^{\text{act}}$ is the signal output power in ARIS, and $P_{\text{DC}}$ is the direct current biasing power.

Fig.~\ref{fig_1} (a) shows the OP for PRIS-NOMA and ARIS-NOMA systems with infinite-precision ADCs (referred to as full-precision ADCs). 
It demonstrates a significant performance disparity in OP between the proposed ARIS-NOMA and PRIS-NOMA systems. Specifically, the ARIS-NOMA system consistently achieves a lower OP compared to the PRIS-NOMA system.
This disparity arises because PRIS experiences the double-fading effect, whereas ARIS enhances the received signal strength at users by amplifying the attenuated signal from the BS with a small amount of power. Fig.~\ref{fig_1} (b) illustrates that outage performance degrades as ADC precision decreases, primarily due to the increased quantization errors.
Fig.~\ref{fig_1} illustrates that the ARIS-NOMA scheme achieves superior outage performance compared to ARIS-OMA, highlighting its improved fairness and efficiency in simultaneously serving multiple users.
Furthermore, Fig.~\ref{fig_1} reveals that ipSIC adversely affects the OP in both PRIS-NOMA and ARIS-NOMA systems, as evidenced by the convergence to error floors in the high SNR regime. This observation suggests that residual interference resulting from ipSIC constitutes a critical limiting factor in the high SNR regions.

Fig.~\ref{fig_2} (a) shows that the presence of RHIs degrades the outage performance of both quantized ARIS-NOMA and PRIS-NOMA systems. In particular, the 1-bit ARIS-NOMA system exhibits superior outage performance compared to its 1-bit PRIS-NOMA counterpart.
Fig.~\ref{fig_2} (b) further elucidates that ARIS-NOMA achieves superior outage performance while necessitating fewer reflecting elements than PRIS-NOMA, thereby substantially reducing the hardware costs and complexity of RIS deployment. Additionally, Fig.~\ref{fig_2} indicates that the negative impacts of low-precision ADCs can be mitigated by optimizing the transmit power and increasing $M$.


Fig.~\ref{fig_3} (a) shows that the 1-bit quantized ARIS-NOMA system reaches its throughput ceiling earlier than the PRIS-NOMA system. This indicates that the ARIS-NOMA system provides higher spectral efficiency and a greater capacity for serving multiple users compared to the PRIS-NOMA system.
Moreover, RHIs degrade system throughput, and residual interference from ipSIC further constrains the throughput ceiling.
As depicted in Fig.~\ref{fig_3} (b), the throughput increases with $M$ and approaches the target rate in the high SNR region. 
Notably, user 3 in ARIS-NOMA achieves comparable throughput with only $10$ reflecting elements, compared to 15 required by PRIS-NOMA. 
This highlights the practicality of quantized ARIS-NOMA systems with low-precision ADCs, especially in space-constrained deployments where minimizing the number of RIS elements is crucial for efficient system implementation.
 \vspace{-9pt} 
\section{Conclusions}
This study analyzed the performance of quantized ARIS-NOMA and PRIS-NOMA systems under RHIs and ipSIC, deriving and validating analytical approximations and asymptotic expressions for OP, diversity order, and throughput. Simulation results demonstrate that RHIs and ipSIC significantly degrade outage performance.
The quantized ARIS-NOMA outperforms PRIS-NOMA in terms of outage performance and throughput, while requiring lower transmit power and fewer reflecting elements.
These advantages highlight the quantized ARIS-NOMA as a highly effective solution for next-generation wireless networks in resource-constrained environments. 
Future research could focus on developing more realistic ipSIC models and practical channels, and on designing joint power allocation and phase-shift optimization strategies for ARIS-NOMA systems under RHIs and residual interference, to enhance their real-time adaptability in dynamic wireless environments.
 \vspace{-12pt} 
{\appendix
\section*{Appendix~A: Proof of the Theorem 1} \label{Appendix:A}
\renewcommand{\theequation}{A.\arabic{equation}}
\setcounter{equation}{0}
Substituting \eqref{e_4} and \eqref{e_5} into \eqref{e_8}, we obtain $P_{act,k}^{ipSIC}$ as
\begin{align}
\label{e_10} \nonumber
&P_{act,k}^{ipSIC}= 1 - \Pr \left( {\sum\limits_{m = 1}^M {\left| {{g_{k,m}}} \right|} \left| {{h_{sr,m}}} \right|  > } \right.\\ \nonumber
&\left. {\sqrt {\varphi _{}^{act*} d_{sr}^\alpha\left( { {\beta ^2}{N_r}{\left\| {{\bf{g}}_k^H{\bf{\Phi }}} \right\|^2}{\Delta _1}+ d_k^\alpha \left( {\lambda _q^2 + \epsilon \rho _s^{act}y} \right)} \right)} } \right)\\ \nonumber
&  = \int_0^\infty  {\frac{1}{{{\Omega _I}}}{e^{ - \frac{y}{{{\Omega _I}}}}}} \frac{1}{{\Gamma \left( {{\mu _0} + 1} \right)}}\gamma \left( {{\mu _0} + 1,\frac{1}{{{\varphi _0}}}\sqrt {\varphi _{}^{act*}d_{sr}^\alpha } } \right. \times \\
&\left. {\sqrt {{\beta ^2}M{N_r}{\Delta _1} + d_k^\alpha \left( {\lambda _q^2 +\epsilon \rho _s^{act}y} \right)} } \right)dy,
\end{align}
where $y= {\left| {{h_I}} \right|^2}$, ${\left\| {{\bf{g}}_k^H{\bf{\Phi }}} \right\|^2}$ denotes the sum of $M$ i.i.d. exponential random variable, and subjects to Gamma distribution with
the parameters of $\left( {M,1} \right)$. The power of the thermal noise
 term introduced by reflection amplifiers in \eqref{e_1} is expressed as $d_{k}^{ - \alpha }{\beta ^2}M{N_r}$. 
 
 Define $z = {y \mathord{\left/
 {\vphantom {y {{\Omega _I}}}} \right.
 \kern-\nulldelimiterspace} {{\Omega _I}}}$, $P_{act,k}^{ipSIC}$ is calculated as
 \begin{align}
\label{e_11} \nonumber
&P_{act,k}^{ipSIC} = \int_0^\infty  {{e^{ - z}}} \frac{1}{{\Gamma \left( {{\mu _0} + 1} \right)}}\gamma \left( {{\mu _0} + 1,\frac{1}{{{\varphi _0}}}\sqrt {\varphi _{}^{act*}d_{sr}^\alpha }  \times } \right.\\
&\left. {\sqrt {{\beta ^2}M{N_r}{\Delta _1} + d_k^\alpha \left( {\lambda _q^2 + \rho _s^{act}{\Omega _I}z} \right)} } \right)dz.
\end{align}
 Thus, applying the Gauss-Laguerre integration method \cite{paper_16} to (\ref{e_11}), $P_{act,k}^{ipSIC}$ can be obtained, completing the proof.
  \vspace{-3pt} 
\section*{Appendix~B: Proof of the Theorem 2} 
\label{Appendix:B}
\renewcommand{\theequation}{B.\arabic{equation}}
\setcounter{equation}{0}
According to \eqref{e_8}, $P_{pas,k}^{ipSIC}$ is expressed as
\begin{align}
\label{e_13} \nonumber
&P_{pas,k}^{ipSIC} = 1 - \Pr \left( {\sum\limits_{m = 1}^M  \left| {{g_{k,m}}} \right|\left| {{h_{sr,m}}} \right|} \right. \ge \\ \nonumber
&\left. {\sqrt {\varphi _{}^{pas*}d_{sr}^\alpha d_k^\alpha \left( {\lambda _q^2 + \rho _s^{pas}{{\left| {{h_I}} \right|}^2}} \right)} } \right)\\ \nonumber
& = \int_0^\infty  {\frac{1}{{{\Omega _I}}}{e^{ - \frac{y}{{{\Omega _I}}}}}} \frac{1}{{\Gamma \left( {{\mu _0} + 1} \right)}} \times \\
&\gamma \left( {{\mu _0} + 1,\frac{1}{{{\varphi _0}}}\sqrt {\varphi _{}^{pas*}d_{sr}^\alpha d_k^\alpha \left( {\lambda _q^2 + \epsilon\rho _s^{pas}y} \right)} } \right)dy.
\end{align}
Defining $z = {y \mathord{\left/
 {\vphantom {y {{\Omega _I}}}} \right.
 \kern-\nulldelimiterspace} {{\Omega _I}}}$ and applying the Gauss-Laguerre integration, $P_{pas,k}^{ipSIC}$ can be obtained. The proof is completed.}

 \vspace{-2pt} 
\balance

%
%
%
%

\vfill

\end{document}